\documentclass[aps,prb,twocolumn,showpacs,superscriptaddress]{revtex4-2}
\usepackage{graphicx}
\usepackage{amsmath}
\usepackage{multirow}
\usepackage{tabularx}
\usepackage{color}
\usepackage{xcolor}
\usepackage{ulem}
\usepackage[colorlinks,linkcolor=blue,urlcolor=blue,anchorcolor=blue,citecolor=blue]{hyperref}

\makeatletter

\newcommand{\Rmnum}[1]{\expandafter@slowromancap\romannumeral #1@}
\newcommand{\Keywords}[1]{\par\noindent\textbf{Keywords:} #1}
\makeatother
\begin{document}
\preprint{APS/123-QED}
\title{High-T$_C$ superconductivity in La$_3$Ni$_2$O$_7$ based on the bilayer two-orbital t-J model}
\date{\today}

\author{Zhihui Luo}
\thanks{These authors contributed equally to this work}

\author{Biao Lv}
\thanks{These authors contributed equally to this work}
\author{Meng Wang}
\author{W\'ei W\'u}
\email{wuwei69@mail.sysu.edu.cn}
\author{Dao-Xin Yao}
\email{yaodaox@mail.sysu.edu.cn}
\affiliation{Center for Neutron Science and Technology, Guangdong Provincial Key Laboratory of Magnetoelectric Physics and Devices, State Key Laboratory of Optoelectronic Materials and Technologies, School of Physics, Sun Yat-Sen University, Guangzhou, 510275, China}

\begin{abstract}
The recently discovered high-$T_{\mathrm{c}}$ superconductor La$_3$Ni$_2$O$_7$ has sparked renewed interest in the unconventional superconductivity.  Here we study superconductivity in pressurized La$_3$Ni$_2$O$_7$ based on a bilayer two-orbital $t-J$ model, using the renormalized mean-field theory. Our results reveal a robust $s^\pm-$wave pairing driven by the inter-layer $d_{z^2}$ magnetic coupling, which exhibits a transition temperature within the same order of magnitude as the experimentally observed $T_{\mathrm{c}} \sim 80$\ K. We establish a comprehensive superconducting phase diagram in the doping plane. Notably, the La$_3$Ni$_2$O$_7$ under pressure is found situated roughly in the optimal doping regime of the phase diagram.  
When the $d_{x^2-y^2}$ orbital becomes close to half-filling, $d-$wave and $d+is$ pairing can emerge from the system.  We discuss the interplay between Fermi surface topology and  different  pairing symmetries.
The stability of the $s^\pm-$wave pairing against Hund's coupling and other magnetic exchange couplings is discussed.\\
\end{abstract}

\maketitle

\section{Introduction}
Understanding high transition temperature ($T_{\mathrm{c}}$) superconductivity remains one of the greatest challenges in the condensed matter physics.
For cuprate superconductors \cite{bednorz1986possible,RevModPhys.78.17,keimer2015quantum}, the fundamental  mechanism of the $d-$wave pairing is believed to primarily lie in the $d_{x^2-y^2}$ orbital,  with strong Coulomb repulsion between electrons playing a  crucial role\cite{RevModPhys.78.17}. This form of superconductivity is usually referred to as unconventional, distinguishing it from the traditional Bardeen-Cooper-Schrieffer (BCS) type of superconductivity. Another prominent example of the unconventional superconductivity is found in the iron-based superconductors \cite{ren2008superconductivity,chen2008superconductivity,chenGF2008superconductivity,ren2019superconductivity,paglione2010high}, where multiple $d-$orbitals participate in the pairing process. The discovery of superconductivity in infinite-layer Nd$_{1-x}$Sr$_{x}$NiO$_2$ thin films has sparked widespread interest due to cuprate-like Ni$^{1+}(3d^{9})$\cite{li2019superconductivity,PhysRevLett.125.027001,PhysRevLett.125.077003,kitatani2020nickelate,PhysRevB.101.081110,kang2023infinite,PhysRevB.101.060504,gu2020single}. Most recently, a Ruddlesden-Popper nickelate  superconductor La$_3$Ni$_2$O$_7$ is found with a $T_\mathrm{c} \approx 80$\ K \cite{sun2023signatures,hou2023emergence,zhang2023hightemperature} under moderate pressures. 
On one hand, La$_3$Ni$_2$O$_7$ shares similarities with cuprates, as both feature the NiO$_2$/CuO$_2$ plane containing a key $d_{x^2-y^2}$ orbital at Fermi level. On the other hand, La$_3$Ni$_2$O$_7$ differs from cuprates as its apical O-$p_z$ and  Ni-$d_{z^2}$ orbitals play a significant role in its low-energy physics~\cite{sun2023signatures,liu2023evidence}.  Given this context, one would ask that whether the underlying  pairing mechanism of La$_3$Ni$_2$O$_7$ resembles that of cuprates?  How does it  differ from the extensively studied cuprates? From a theoretical perspective, a first step in addressing these questions is to resolve the roles played by the $d_{x^2-y^2}$ and $d_{z^2}$ orbitals  in pairing, and  to construct a superconducting phase diagram of the relevant physical models.

Electronic structure studies \cite{luo2023bilayer,lechermann23,zhang2023electronic,sakakibara2023possible,shilenko2023correlated,jiang2023pressure,geisler2024structural} and optical experimental probe \cite{liu2023electronic} suggest that in  pressurized La$_3$Ni$_2$O$_7$,  Ni-$d_{z^2}$ orbital is involved in Fermi energy due to strong inter-layer coupling via apical oxygen. Through hybridization with the in-plane oxygen $p-$orbital, $d_{z^2}$ orbitals can mix with $d_{x^2-y^2}$ orbitals, resulting in a three-pocket Fermi surface structure\cite{luo2023bilayer}. This Fermi surface geometry differs from that of cuprates, which may lead to pairing symmetry and effective pairing ``glue'' distinct from that of the $d-$wave superconductivity in cuprates, which arises upon doping a single $d_{x^2-y^2}$ orbital. 
Another crucial consideration is the electron occupancy. La$_3$Ni$_2$O$_7$ typically exhibits a nominal valence configuration of  $d^{7.5}$ \cite{sun2023signatures,pardo2011metal}, indicating an average of  2.5 holes in the active $e_g$ sub-shell. While previous studies have shown variations in the computed electron densities of the two $e_g$ orbitals\cite{wú2023charge,yang2023minimal,christiansson2023correlated,qin2023hightc,jiang2023high}, it can be in general viewed as a multi-orbital system comprising a heavily hole-doped  $d_{x^2-y^2}$ orbital ( with a hole density $\sim 1.5$), and a near half-filled  $d_{z^2}$ orbital ( with a hole density  $\sim 1.0$). Regarding superexchange couplings, investigation based on cluster dynamical mean-field theory \cite{wú2023charge} have pointed out that the exchange coupling between two inter-layer $d_{z^2}$ orbitals $J_{\bot}$ may be substantially larger, by a factor of at least $\sim$2 ,  than the intra-layer $d_{x^2-y^2}$ exchange coupling $J_{||}$, with the latter estimated to be comparable to its cuprate counterpart \cite{wú2023charge}. Such a significant $J_\bot$ is  likely responsible for the high transition $T_\mathrm{c}$ in La$_3$Ni$_2$O$_7$ \cite{wú2023charge,yang2023minimal,shen2023effective,yang2023possible,lu2023interlayer,qu2023bilayer,liu2023spmwave}.

In this paper, we systematically investigate superconductivity in the bilayer two orbital $t-J$ model, which serves as a prototype for the low-energy physics of pressurized La$_3$Ni$_2$O$_7$,  using the renormalized mean-field theory (RMFT) \cite{FCZhang_1988,anderson2004,wang2006unrestricted}.
The RMFT approach is a concrete implementation of Anderson’s resonating valence bond (RVB) concept of the unconventional superconductivity~\cite{anderson1987resonating}.  Following the Gutzwiller scheme, the renormalization effects from strong electron correlations are accounted on different levels by incorporating the doping-dependent renormalization factors $g_t,g_J$ \cite{FCZhang_1988} in RMFT.  The mean-field decomposition of the magnetic exchange couplings $J$, then enables exploring of the BCS pairing instabilities within the system.
Despite its straightforward formulations, RMFT has demonstrated its capability of capturing various aspects of cuprate superconductors, including superfluid density, the dome-shaped doping dependence of $T_\mathrm{c}$, and  pseudogap phenomena~\cite{anderson2004}. 
In RMFT, the superconducting order parameter $g_t\Delta$ involves two competing energy scales: the phase coherence energy scale associated with $g_t$, and the pairing amplitude represented by $\Delta$, each exhibiting distinct doping dependencies~\cite{FCZhang_1988}. In our study, we observe that the outcome of this competition positions La$_3$Ni$_2$O$_7$ within the optimal doping regime in the superconducting phase diagram. Our calculation suggests a $T_\mathrm{c}$ comparable to experimental observations, indicating the relevance of our considerations to the La$_3$Ni$_2$O$_7$ superconductor. Furthermore, we elucidate various possible pairing symmetries across a broad doping range.

  The remainder of paper is organized as follows. In Sec.~\ref{sec:theory}, we present the physical model and  describe the RMFT method. In Sec.~\ref{sec:results}, we present our results including $T_\mathrm{c}$ for the pristine compound, the doping phase diagram,  and the Fermi surface.  Additionally, the impact of the strength of superexchanges and Hund's coupling is discussed at the end of the section. In Sec.~\ref{sec:discussion} we discuss the stability  of the  $s^\pm-$pairing. Finally,  more details  can be found in the Supplementary Information.

\section{RESULTS}
\label{sec:results}
Now we present the RMFT result on the superconducting instabilities of the bilayer two-orbital $t-J$ model. In particular, we provide detailed investigations in the parameter regime that most relevant to the $\mathrm{La_3Ni_2O_7}$ system. The impacts of several key factors including temperature $T$, doping levels of $d_{x^2-y^2}$, and $d_{z^2}$ orbitals:  $p_x,p_z$, as well as the geometry of Fermi surface are analyzed while eyeing on the evolution of the superconducting order parameter $g_t^{\alpha\beta}|\Delta_{\ell}^{\alpha\beta}|$.
The definition of $g_t^{\alpha\beta}|\Delta_{\ell}^{\alpha\beta}|$ and RMFT formalism can be found in Methods.  
For brevity, we adopt the shorthand notation $g_t^\nu|\Delta^{\nu}_\ell|$ in the following discussion to represent $g_t^{\alpha\beta}|\Delta_{\ell}^{\alpha\beta}|$, where $\ell=d,s^\pm$, denoting pairing symmetries, and $\nu=||x,||z,\bot z$ denoting intra-layer $d_{x^2-y^2}$ [$(\alpha, \beta) = (x_1, x_1)$], intra-layer $d_{z^2}$ [$(\alpha, \beta) = (z_1, z_1)$] and inter-layer $d_{z^2}$  [$(\alpha, \beta) = (z_1, z_2)$] pairing.  Without loss of generality,  we adopt typical values of $J_\bot =2J_{||}=0.18$\ eV taken from Ref.~\cite{wú2023charge} throughout the paper unless otherwise specified.

\subsection{\label{subsec:Tc}$T_\mathrm{c}$ for pristine compound}
\begin{figure}
\includegraphics[scale=0.7,trim=24 0 0 0]{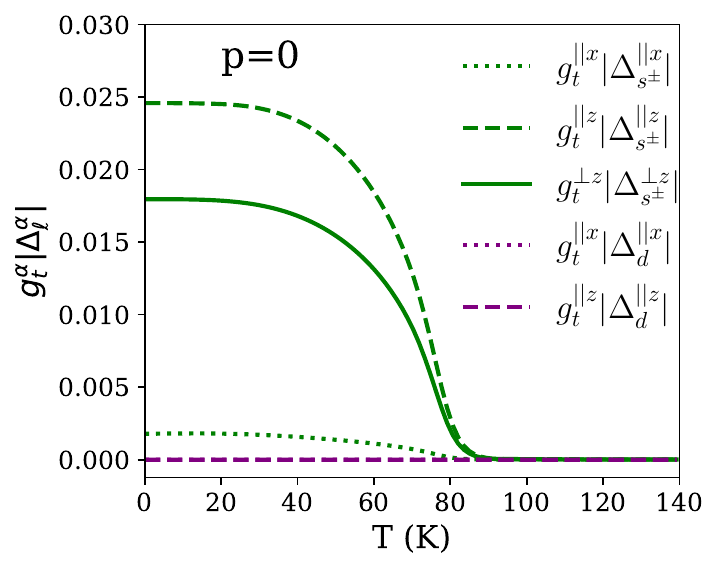}\caption{\label{fig:T}Superconducting order parameters $g_t^{\nu}|\Delta^\nu_{\ell}|$ as a function of temperature $T$. Note that the two purple lines overlap with each other with zero magnitude over the whole temperature range. Since here the doping is fixed as $p=0$ ($n_x = 0.665 , n_z = 0.835$), $g_t^{\nu}$ is constant according to Eq.~(\ref{eq:gt}-\ref{eq:mfij}). }
\end{figure}

We first present the RMFT calculated superconducting transition temperature $T_\mathrm{c}$ at $\mu=0$ that corresponds to pristine La$_3$Ni$_2$O$_7$ under pressure, which is to be dubbed as the pristine compound (PC) case hereafter. For the PC case, we have $\mu=0, n_x = 0.665 , n_z = 0.835$, and  $n=n_x+n_z = 1.5$ \cite{luo2023bilayer}. 
In  Fig.~\ref{fig:T}, the superconducting order parameter  $g_t^{\nu}|\Delta^\nu|$ is plotted as a function of $T$, which clearly demonstrate two dominant branches of the pairing fields at small $T$: the intra-layer $d_{z^2}$ pairing (dashed line) and the inter-layer $d_{z^2}$ pairing (solid line), forming the  $s^{\pm}-$wave pairing of the system. This result is in agreement with several other theoretical studies 
\cite{liu2023spmwave,yang2023possible,gu2023effective,qu2023bilayer,zhang2023structural,tian2023correlation}. As increasing $T$, the order parameters decrease in a mean-field manner and eventually drop to zero at around 80\ K. The computed value of $T_\mathrm{c}$ is somehow coincides with the experiment \cite{sun2023signatures}, highlighting that the various energy scales under our consideration can effectively capture the major physics of the realistic compound. 
However, we would like to stress that the superconducting $T_\mathrm{c}$ from RMFT in fact dependent on the value of $J$ essentially in a BCS manner. Hence it can be sensitive to the strength of the superexchange couplings.  The coincidence between the experimental and RMFT value of $T_\mathrm{c}$ should not be taken as the outcome of RMFT capturing  La$_3$Ni$_2$O$_7$ superconductivity in a quantitatively correct way.  We note that the $s^{\pm}-$wave pairing  has also finite  $d_{x^2-y^2}$ orbital component, as shown by the dotted lines in 
Fig.~\ref{fig:T}, despite that its order parameter are much smaller than that of the $d_{z^2}$ orbitals. The $d-$wave order parameters (purple), on the other hand, are fully suppressed, suggesting that the  $d+is-$wave pairing instability may be ruled out in our model for La$_3$Ni$_2$O$_7$. 
It is worthy noting that RMFT is originally formulated at zero temperature~\cite{FCZhang_1988}. 
 In Fig.~\ref{fig:T} we solve the RMFT equations in the finite temperature regime~\cite{wang2010finite}.  This extension captures finite temperature effects on the mean-field parameters $\Delta$ and $\chi$, while neglecting thermal effects on the renormalization factors $g_t, G_{J_r}$. We argue that the latter should have insignificant impacts on determining $T_\mathrm{c}$,  given the hole concentrations in Fig.~\ref{fig:T},  ($n^{h}_x =1-n_x\approx 0.335 , n^h_z = 1-n_z \approx 0.165$) are fairly large. Detailed discussions on this matter can be found in the Supplementary Note 2.

\begin{figure}[ht]
	\centering
	\includegraphics[width=1\columnwidth]{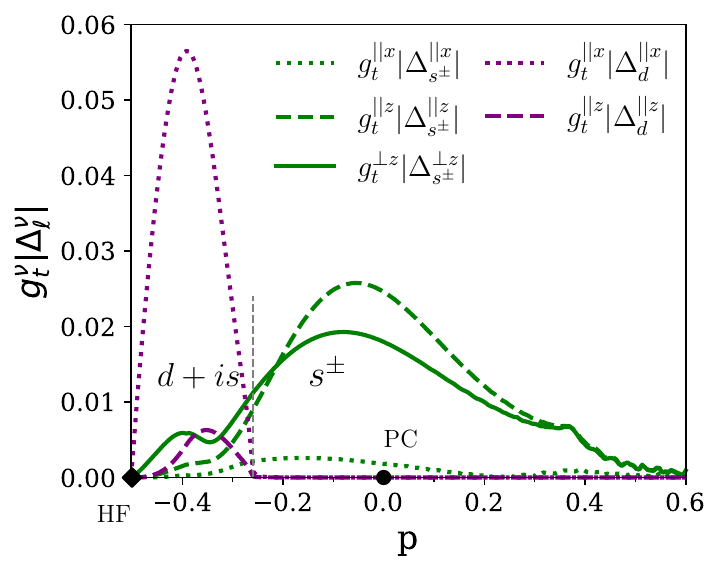}
	\caption{Superconducting order parameter $g_t^{\nu}|\Delta^\nu_{\ell}|$ as a function of doping  $p$. $p>0$ represents hole doping the pristine compound case (where $n_x = 0.665 , n_z = 0.835$),  and $p<0$ for electron doping. Black dot at $p=0$ indicates the pristine compound (PC). We vary $p$ while a fixed ratio of $p_x/p_z=2.048$ is maintained. The two $e_g$ orbitals reach half-filling (HF) ($n_x = 1.0 , n_z = 1.0$) at $p=-0.5$ , as indicated by the diamond symbol.}
	\label{fig:p}
\end{figure}

\subsection{\label{subsec:doping}Doping evolution}
Now we focus on the doping dependence of the superconducting order parameter $g_t^{\nu}|\Delta^\nu_{\ell}|$. In the following,  $p>0$ means doping the pristine compound ( where $n_x = 0.665 , n_z = 0.835$) with holes, and $p<0$ denotes electron doping. While varying the doping level  $p$ of the system,  we maintain a fixed ratio between the doping levels of the two $e_g $ orbitals, namely, $p_x/p_z=2.048$ is fixed, such that both $e_g$ orbitals are half-filling (HF), \textit{i.e.,} $n_x=1, n_z=1$ at  $p=-0.5$. 
 From Fig.~\ref{fig:p}, one learns that the $s^{\pm}-$wave pairings (green) are quite robust over a wide range of doping $p$. The maxima of $g_t^{\nu}|\Delta_{s^\pm}^{\nu}|$ are located at $p \approx -0.04$ which is very close to $p=0$ for PC. This indicates that, interestingly, the La$_3$Ni$_2$O$_7$ under pressure corresponds to roughly the optimal doping in our superconducting phase diagram.  At extremely large electron dopings ($p<-0.25$), $d-$wave pairing can build up  which also exhibits a predominant superconducting dome (purple dotted line). In this doping regime, small $s^\pm-$wave components of the superconducting order parameter are found to coexist with the $d-$wave components,  indicating the emergence of $d+is-$wave pairing.
At half-filling in Fig.~\ref{fig:p} ($p=-0.5$), all pairing channels are fully suppressed due to the vanishing renormalization factors $g_t^{\nu}\rightarrow0$, reflecting the Mott insulting nature at half-filling \cite{wú2023charge}.

It is worth noting that, as a general prescription of the mean-field approaches, different $J_r$ terms in Eq.~\ref{eq:hjmf} can be decomposed into different corresponding pairing bonds $\Delta_{\delta}$,  such as  $\Delta^{\bot z}_{d/s^\pm}$ from decomposing $J_\bot $,   and  $\Delta^{||x}_{d/s^\pm}$ from decomposing $J_{||}$ . 
Note that in our calculation, the pairing components $\Delta^{||z}_{d/s^\pm}$ (dashed lines in Fig.~\ref{fig:p}) that represent the intra-layer pairing of $d_{z^2}$ orbital, do not have a corresponding $J$ term in the Hamiltonian. Their values are  not determined by the competition between  $\Delta^{||z}_{d/s^\pm}$ and $|\Delta^{||z}_{ d/s^\pm}|^2$ terms in minimizing the free-energy. Instead, it should be interpreted as the pairing instability induced by the pre-existing inter-layer $d_{z^2}$ pairing. Indeed, as shown in Fig.~\ref{fig:p},  $\Delta^{||z}_{s^\pm}$ displays a doping $p-$dependence  similar to that of $\Delta^{\perp z}_{s^\pm}$. 
Finally, we notice that a small tip appears at hole doping $p\sim 0.4$, which can be attributed to the van Hove singularity associated with the $\beta-$sheet of the Fermi surface, see also Section~\ref{subsec:fs}.

\subsection{\label{subsec:phasediagram}Doping phase diagram }
\begin{figure}[ht]
	\centering
	\includegraphics[width=1\columnwidth]{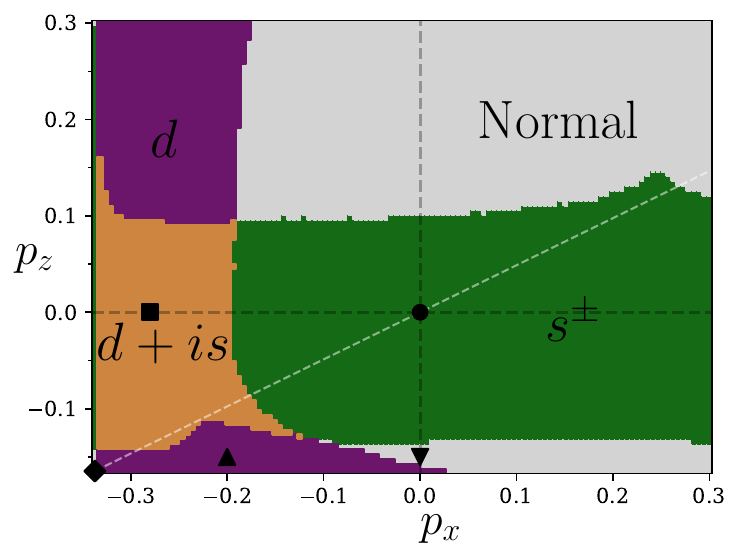}
	\caption{Pairing phase diagram constructed by varying doping levels $p_x$ and $p_z$. $p_\alpha>0$ indicates hole doping and $p_\alpha<0$ denotes electron doping relative to the pristine compound case ($n_x = 0.665 , n_z = 0.835$). The dashed white line presents the $p_x-p_z$ trajectory along which     Fig.~\ref{fig:p} is plotted. The black symbols mark the sets of ($p_x,p_z$) that are further discussed in Fig.~\ref{fig:fs}. The superexchange couplings applied are $J_\bot =2J_{||}=0.18$.}
	\label{fig:phasediagram}
\end{figure}
\begin{figure}[ht]
	\centering
	\includegraphics[width=1\columnwidth]{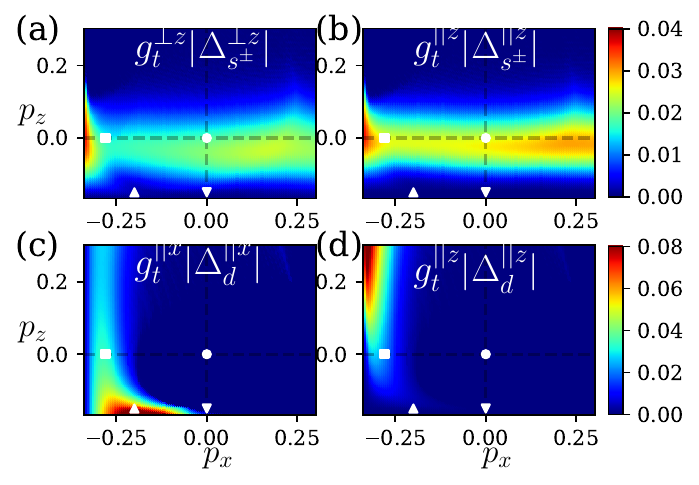}
	\caption{Superconducting order parameters $g_t^{\nu}|\Delta^\nu_{\ell}|$ with varying doping levels $p_x$ and $p_z$. Again, here $(p_x=0, p_z=0)$ corresponds to $(n_x = 0.665 , n_z = 0.835)$.
  Symbols denote typical dopings to be analyzed in Fig.~\ref{fig:fs}.}
	\label{fig:phasediagram2}
\end{figure}

To gain further insights into the RMFT result of La$_3$Ni$_2$O$_7$ system, we obtain a phase diagram in the  $p_x -p_z$ doping plane where now the two dopings are independent variables. As shown in Fig.~\ref{fig:phasediagram}, RMFT reveals that $d$, $s^\pm$, and $d+is$ pairing symmetries, as well as the normal state occur in different doping regimes.  Here a dashed white line indicates the $p_x - p_z $ trajectory along which Fig.~\ref{fig:p} is plotted. The black symbols label out sets of ($p_x,p_z$) parameters  on which the result will be further discussed in Fig.~\ref{fig:fs}. The major feature of Fig.~\ref{fig:phasediagram} is that the $d-$wave and $s^\pm-$wave pairings span roughly a vertical and a horizontal stripe  respectively in the phase diagram. In other words, $s^\pm-$wave pairing  (green) dominates the regime where $ -0.1 \lesssim p_z \lesssim  0.1$ ($0.735\lesssim n_z \lesssim 0.935$), and it is insensitive to the value of $p_x$. Likewise, the $d-$wave pairing  (purple) prevails in the doping range of $ -0.335  \lesssim p_x \lesssim  -0.2$ (0.465 $\lesssim n_x \lesssim$ 1.0), and it is in general less  sensitive to the value of $p_z$.  
As a result, the $d+is-$wave pairing  (orange) naturally emerges at the place where the two stripes overlap. 
In order to have a better understanding of this phase diagram, we show the magnitudes of the four major pairing bond $g_t^{\nu}|\Delta^\nu_{\ell}|$ in Fig.~\ref{fig:phasediagram2} , from which one sees that for $s^\pm-$wave pairing, the pairing tendencies of $\Delta_{s^\pm}^{\bot z}$ (Fig.~\ref{fig:phasediagram2}a) and $\Delta^{||z}_{s^\pm}$ (Fig.~\ref{fig:phasediagram2}b) show similar pattern in the $p_x - p_z$ plane, in consistence with Fig.~\ref{fig:p}. For the $d-$wave pairing, the situation is however different. For intra-layer pairing, $g^{||x}_t|\Delta^{||x}_{d}|$ is enhanced when $d_{x^2-y^2}$ is heavily electron doped ($p_x \approx -0.25$), and $d_{z^2}$ becomes half-filling ($p_z \sim -0.2$). This is because  the large electron doping drives the $\gamma-$pocket of $d_{z^2}$ orbital centered at $M-$point descends into the Fermi sea. Such that the system becomes effectively a single band system of the active $d_{x^2-y^2}$ orbital. Hence the dominant $d-$wave pairing of the single-band $t-J$ model is recovered for $d_{x^2-y^2}$ orbital in this limit, mimicking the physics of cuprates. On the other hand, the $d_{z^2}$ component of the $d-$wave pairing $g^{||z}_t|\Delta^{||z}_{d}|$ 
is enhanced when $p_z > 0.2$, as shown in Fig.~\ref{fig:phasediagram2}d. This can be understood considering the fact that since intra-layer exchange of $d_{z^2}$ orbital $J_{||z}=0$ in our study, the $d-$wave instability driven by  $J_{||}$ of the $d_{x^2-y^2}$ orbital is less sensitive to the details of $d_{z^2}$ orbital. Hence,  the order parameter $g^{||z}_t | \Delta^{||z}_{d} |$  increasing  with $p_z$ shown in  Fig.~\ref{fig:phasediagram2}d can bee seen vastly as a result of a growing $g^{||z}_t$ with decreasing $n_z$ according to  Eq.~(\ref{eq:gt}-\ref{eq:mfij}).
Finally, it is interesting to note that although  $J_{\bot} =  2 J_{||}$ is used in our study, 
Fig.~\ref{fig:phasediagram2} shows that the the maximal value of $d-$wave superconducting order parameter are roughly two times larger than that of the $s^\pm-$wave pairing. This is expected since the vertical exchange coupling  $J_{\bot}$ has a smaller coordination number $z=1$, compared to its the in-plane counterpart $J_{||}$, where $z=4$.



\subsection{\label{subsec:fs}Fermi surfaces}
\begin{figure}[h]
	\centering
	\includegraphics[width=0.8\columnwidth]{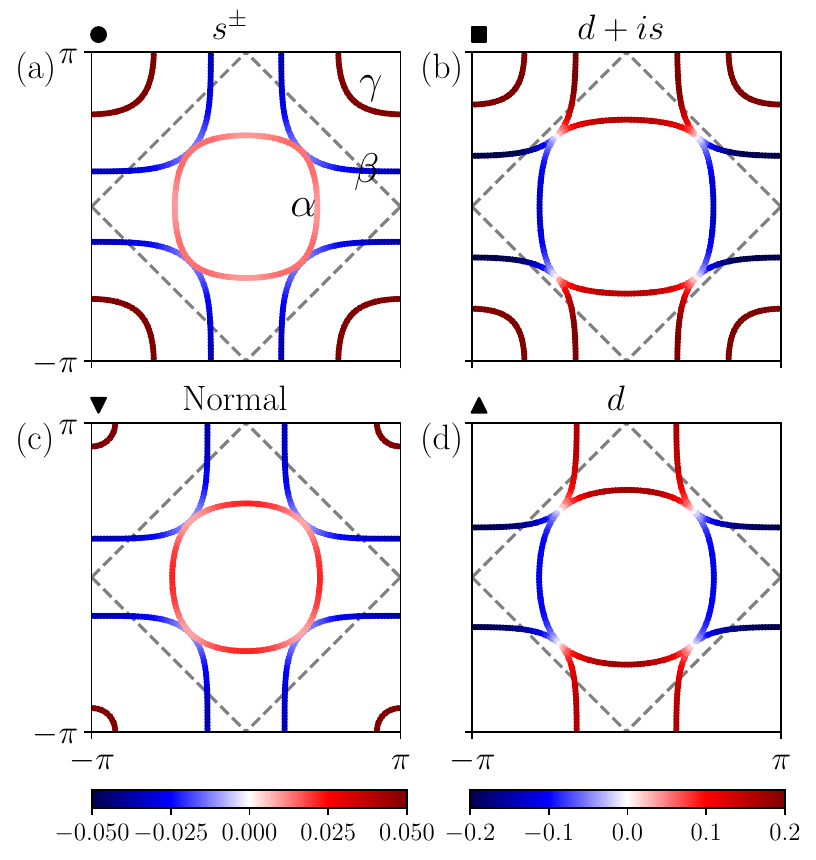}
	\caption{The pairing gap function $\Delta^{\nu}_{\ell}$ projected onto the Fermi surface for a  few sets of ($p_x,p_z$). Dot: $(p_x=0,p_z=0)$, square:  $(p_x=-0.28,p_z=0)$, upper triangle: $(p_x=-0.2,p_z=-0.15)$, and lower  triangle: $(p_x=0,p_z=-0.15)$.  See also symbols indicated in Fig.~\ref{fig:phasediagram}  and \ref{fig:phasediagram2}.  Note that the gap magnitudes are directly associated with  $\Delta^{\nu}_\ell$ instead of the superconducting order parameter $g_t^{\nu}|\Delta_\ell^{\nu}|$, which is why (c) shows nonvanishing gap for a normal state.  The dashed lines indicate the folded Brillouin zone (FBZ) of the two inplane Ni. }
	\label{fig:fs}
\end{figure}

In Fig.~\ref{fig:fs} we display the paring gap function $\Delta^\nu_\ell$ projected onto the Fermi surfaces for four typical sets of dopings $(p_x, p_z)$ with each characterizing one type of pairing symmetries (Fig.~\ref{fig:fs}a, Fig.~\ref{fig:fs}b, Fig.~\ref{fig:fs}d) or the one with vanishing superconducting order parameter (Fig.~\ref{fig:fs}c). 
Fig.~\ref{fig:fs}a shows the FS of the PC case with $s^\pm$ pairing, which has  three sheets of FS with one around $\Gamma$ point and two around $M$ point \cite{luo2023bilayer}. 
Also, the sign structure is consistent with that in Refs.~\cite{liu2023spmwave,yang2023possible,zhang2023structural}. The key feature here is that the sign of the pairing gap is the same within the $\alpha, \gamma$ bands,  but it  reverses between $\alpha/\gamma$ and $\beta$ pockets. This is because the former two components come from the bonding state while the latter is from the anti-bonding state.
Decreasing $p_x$ from the PC can drives the $d_{x^2-y^2}$ orbital closer to half-filling. As shown in Fig.~\ref{fig:fs}b,  the $\alpha,\beta-$sheet of FS as a whole is also driven closer to the folded Brillouin zone (FBZ) edge (dashed lines), which is accompanied by the  the pairing symmetry evolving from $s^\pm$ to $d+is-$wave. 
In this case, we see that the $\alpha,\beta$ pockets exhibit the $d-$wave sign structure while the $\gamma$ pocket maintains  the $s$-wave symmetry, and, its profile is less affected by the changing of $p_x$.
The occurrence of the $d-$wave pairing at this doping level unambiguously signals the importance of the intra-orbital physics in $d_{x^2-y^2}$ orbital as it approaches half-filling.
Fig.~\ref{fig:fs}c displays that as lowing $p_z$ from the PC case, the $\gamma-$pocket vanishes 
from the Brillouin zone. Consequently, the $s^\pm-$wave order parameter vanishes at $p_z \sim -0.15$.  In this case, similar to PC, no finite $d-$wave order parameter is observed. 
Note that the gap magnitude in Fig.~\ref{fig:fs} is associated with $\Delta^\nu_\ell$ instead of the superconducting order parameter $g_t^\nu|\Delta_\ell^\nu|$, which is why Fig.~\ref{fig:fs}c shows nonvanishing gap for a normal state, see also Supplementary Note 4.
Fig.~\ref{fig:fs}d shows the last case with the $\gamma-$pocket vanishing from Fermi level. As expected, in this case,
 only the $d-$wave pairing with finite $g^{||x}_t|\Delta^{||x}_{d}|$ is found. As discussed above, here the physics of the system can be essentially captured by the single-band $t-J$ model with the presence of only $d_{x^2-y^2}$ orbital.

\subsection{\label{subsec:j}Superexchanges $J$} 

\begin{figure}[ht]
	\centering
	\includegraphics[width=1\columnwidth]{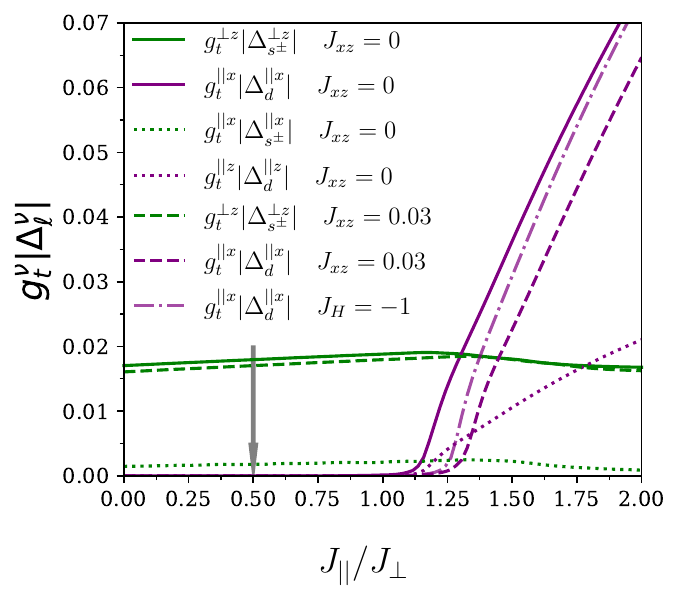}
	\caption{Superconducting order parameters $g_t^{\nu}|\Delta^\nu_{\ell}|$ as a function of $J_{||}/J_\bot$ for the pristine compound. The solid and dotted lines denote cases of $J_{xz}=0$, dashed lines denote that of $J_{xz}=0.03$\ eV, and dash-dotted line denotes the case of $J_H=-1$\ eV.  Here $J_\bot=0.18$\ eV is fixed.  The vertical arrow indicates  $J_{||}/J_\bot=0.5$  which is used in our above calculations.}
	\label{fig:j}
\end{figure}

Finally, we investigate  the influence of the magnitudes of the superexchanges for the pristine compound. In Fig.~\ref{fig:j}, we present $g_t^{\nu}|\Delta^\nu_{\ell}|$ as a function of $J_{||}/J_\bot$. A vertical  arrow indicates the value of $J_{||}/J_\bot$ used in aforementioned calculations, where $s^\pm-$wave is found for PC. As decreasing/increasing $J_{||}/J_\bot$,  $s^\pm-$wave order parameters (green lines) decrease/increase very slightly with $J_{||}/J_\bot$.
When $J_{||}/J_\bot\sim1.1$, $d-$wave (purple solid line) starts to build up at $d_{x^2-y^2}$ orbital. For  La$_3$Ni$_2$O$_7$ under pressure,  this large value of $J_{||}/J_\bot$ is however not very
realistic~\cite{wú2023charge}. Hence, the  $d-$wave pairing instability may can be excluded for the realistic materials in our study.
To check the stability of the superconductivities,  results for $J_{xz}=0.03$ (dashed line) and $J_H=-1$\ {eV} (dash-dotted line) are  shown in Fig.~\ref{fig:j}. As one sees that, although  both  $J_{xz}$  and  $J_H$ act as pair-breaking factors, they do not significantly modify
the result we obtained above. In particular, for the $s^\pm-$pairing, the changes of the order parameter $g_t^{\nu}|\Delta^\nu_{\ell}|$ caused by  $J_{xz}$ (green  dashed line)  and   $J_H$ (not shown here)  are negligible.

\section{\label{sec:discussion}Discussion}
In our study, the RMFT equations are solved in such a way that the pairing fields on different bonds are varied independently, namely, no specific pairing symmetry is presumed  in the self-consistent process. The symmetries of the electron pairing naturally emerge as a result of energy minimization in our calculations, preventing the potential overlooking of pairing symmetries. Regarding the dominant $s^{\pm}$ pairing for the pristine compound at $p_x ,p_z$ = 0, we note that even when only $J_{\perp}$ is considered (with $J_{||}=0,J_{xz} =0 $), $\Delta^{||x}$ is finite although much smaller than  $\Delta^{||z}$ and   $\Delta^{\perp z}$. Given the smaller effective mass of the $d_{x^2-y^2}$ orbital compared to the $d_{z^2}$ orbital, it may still contribute significantly to the superfluid density in the superconducting state of the system. The dual effects of the Hund's coupling $J_H$, namely, the alignment of the on-site spins of the $d_{x^2-y^2}$ and $d_{z^2}$ orbitals, and the enhancement of $J_{||}$ and $J_{xz}$ couplings, show no significant impact on  the dominant $s^{\pm}$ pairing in our RMFT study, as indicated in Fig.~\ref{fig:j}. However,
we acknowledge that the implications of $J_H$ on superconductivity may be underestimated in the RMFT formalism~\cite{multibandGA1998}.
Additionally, it is important to  note that the superconducting $T_\mathrm{c}$ obtained by RMFT is generally overestimated because both temporal and spatial fluctuations are neglected. This is particular true when considering that the pairing fields originate from the local inter-layer $d_{z^2}$ magnetic couplings,  where phase 
fluctuations can play a more significant role in suppressing $T_\mathrm{c}$ comparing to its single-band $t-J$  model counterpart in cuprate superconductors. Finally, a recent theory work \cite{yang2023minimal} proposes that within the framework of the two-component pairing theory in a composite system, the phase fluctuations can be suppressed by the hybridization effects between $d_{x^2-y^2}$ and $d_{z^2}$ orbitals. Verifying this conjecture is, however, beyond the scope of this work.

Employing the renormalized mean-field theory, we have established a comprehensive superconducting phase diagram for the bilayer two-orbital $t-J$ model. A robust $s^{\pm}-$wave pairing  is found  to exist in the parameter regime relevant to La$_3$Ni$_2$O$_7$ under pressure, which in general corresponds to the optimal doping of the superconducting phase diagram. We have carefully investigated the dependence of the pairing instabilities on doping levels, exchange couplings, and the effects of Hund's coupling. Our study provides significant insights into  the theoretical understanding of the superconductivity La$_3$Ni$_2$O$_7$ under pressure.

\section{Methods}\label{sec:theory}
\subsection{Model}
In the strong-coupling limit, the bilayer two orbital Hubbard model \cite{luo2023bilayer} can be mapped into a $t-J$ model~\cite{wú2023charge}

\begin{align}
\label{eq:t-J model}
\mathcal{H}&=\mathcal{H}_t+\mathcal{H}_J \\
\mathcal{H}_t&=\sum_{ij \alpha\beta\sigma} \nonumber (t_{ij}^{\alpha\beta}-\mu\delta_{ij}\delta_{ \alpha\beta} )c_{i \alpha\sigma}^{\dagger}c_{j\beta\sigma} \\
&+\epsilon_{xz}\sum_{i\sigma}^{\alpha\beta=x_1z_1,x_2z_2}(n_{i\alpha\sigma}-n_{i\beta\sigma}) \nonumber \\
\mathcal{H}_J &=J_\bot \sum_{i} \boldsymbol{S}_{iz_1}\cdot \boldsymbol{S}_{iz_2}+J_{||} \sum_{<ij>}^{ \alpha=x_1,x_2} \boldsymbol{S}_{i \alpha}\cdot \boldsymbol{S}_{j \alpha}\nonumber\\
\quad &+J_{xz} \sum_{<i,j>}^{ \alpha\beta=x_1 z_1, x_2 z_2} \boldsymbol{S}_{i \alpha}\cdot \boldsymbol{S}_{j \beta} +J_H \sum_{i}^{ \alpha\beta=x_1z_1,x_2z_2} \boldsymbol{S}_{i \alpha}\cdot \boldsymbol{S}_{i \beta}, \nonumber
\end{align}
where $\mathcal{H}_t$ is the tight-binding Hamiltonian taken from downfolding the DFT band structure \cite{luo2023bilayer}, which is defined in a basis of $\Psi_{ i\sigma}=(c_{ ix_1\sigma},c_{ iz_1\sigma},c_{ ix_2\sigma},c_{ iz_2\sigma})^T$, with $c_{ i\alpha\sigma}$ representing annihilation of an electron with spin $\sigma$ on $ \alpha=x_1,z_1,x_2,z_2$ orbital  at $i$ site . 
 Here $x_1,x_2$ denote the two Ni-$3d_{x^2-y^2}$ orbitals situated in the double  $\mathrm{NiO_2}$ layer while $z_1, z_2$ correspondingly denote the two Ni-$3d_{z^2}$ orbitals. 
$\mu$ is the  chemical potential  and $\epsilon_{xz}$ is the energy difference between $d_{x^2-y^2}$ and $d_{z^2}$ orbitals. 
The hopping parameters in $\mathcal{H}_t$ used in this work  can  be found in Supplementary Note 1. $\mathcal{H}_J$ is the Heisenberg exchange couplings,  and the spin operator $\boldsymbol{S}_{i \alpha}=\frac{1}{2}\sum_{\mu\nu}c^\dagger_{i \alpha\mu}\boldsymbol{\sigma}_{\mu\nu} c_{i\alpha\nu}$. According to the estimated  antiferromagnetic correlations in $\mathrm{La_3Ni_2O_7}$ \cite{wú2023charge}, there can be three major magnetic exchange couplings  $J_\bot ,J_{||},J_{xz}$, which respectively represents nearest-neighbor inter-layer exchange of $d_{z^2}$ orbital, intra-layer exchange of $d_{x^2-y^2}$ and intra-layer exchange between $d_{x^2-y^2}$ and $d_{z^2}$.
$J_H$ is the intra-atomic Hund's coupling.

\subsection{RMFT Formalism}
To proceed with  RMFT \cite{FCZhang_1988}, we first define the following mean-field parameters
\begin{align}
\chi_{ij}^{\alpha\beta}=\frac{1}{2}\langle c_{i\alpha\uparrow}^{\dagger}c_{j\beta\uparrow}+c_{i\alpha\downarrow}^{\dagger}c_{j\beta\downarrow}\rangle=\langle c_{i\alpha\uparrow}^\dagger c_{j\beta\uparrow} \rangle \label{eq:chi}\\
\Delta_{ij}^{\alpha\beta}=\frac{1}{2}\langle c_{i\alpha\uparrow}^{\dagger}c_{j\beta\downarrow}^{\dagger}-c_{i\alpha\downarrow}^{\dagger}c_{j\beta\uparrow}^{\dagger}\rangle = \langle c_{i\alpha\uparrow}^\dagger c_{j\beta\downarrow}^\dagger \rangle,
\label{eq:Delta}
\end{align}

where $\chi_{ij}^{\alpha\beta}$  and $\Delta_{ij}^{\alpha\beta}$ are particle-hole and particle-particle pairs relating $i\alpha$ and $j\beta$. Here we assume no magnetic ordering. For each type of exchange couplings $J_r$ in Eq.~(\ref{eq:t-J model}), including  the Hund's coupling $J_H$, the mean-field decomposition of $\mathcal{H}_{J}$ introduces  condensations of $\chi$ and $\Delta$  in the corresponding $r-$channel
\begin{align}
\label{eq:hjmf}
H_{J_r}^{\chi}&=-\frac{3}{4}J_r \sum_{<ij>\sigma} (\chi^{\alpha\beta}_{\delta}c^\dagger_{i\alpha\sigma}c_{j\beta\sigma}+h.c. )+\frac{3}{2}J_rN|\chi^{\alpha\beta}_\delta|^2, \\ 
H_{J_r}^{\Delta}&= -\frac{3}{4}J_r\sum_{<ij>\sigma} (\sigma \Delta^{\alpha\beta}_{\delta} c^\dagger_{i \alpha\sigma}c^\dagger_{j\beta \bar{\sigma}}+h.c. )+\frac{3}{2}J_rN|\Delta_\delta^{\alpha\beta}|^2.  \nonumber
\end{align}

Assuming translation symmetry, $\delta=R_{j\beta}-R_{i\alpha}$ denotes different bonds in real-space associated with $J_r$, and $N$ is the  total number of sites of the square lattice.

 We now introduce two renormalization factors \cite{zhangfc2003prl}
\begin{eqnarray}
\label{eq:gt}
G_{t}^{\alpha}=\sqrt{\frac{1-n_{\alpha}}{1-n_{\alpha}/2}},\qquad
G_{J}^{\alpha}=\frac{1}{(1-n_{\alpha}/2)}.
\end{eqnarray}
These two quantities essentially reflect the  renormalization effects by the electrons repulsions on top of the single-particle Hamiltonian in the Gutzwiller approximation \cite{FCZhang_1988,gan2005gossamer,wang2006unrestricted,edegger2007Gutzwiller}, which depend on 
 the orbital occupation number $n_{\alpha}=n_{\alpha\uparrow}+n_{\alpha\downarrow}$.
This eventually leads to the renormalized mean-field Hamiltonian
 \begin{align}
  \label{eq:mfij}
 H^{\rm MF}_t&=\sum_{ij\alpha\beta\sigma} g_t^{\alpha\beta}  t_{ij}^{\alpha\beta}c_{i\alpha\sigma}^\dagger c_{j\beta\sigma},\\ 
 H_J^{{\rm MF,\chi/\Delta}}
 &=\sum_{r}G_{J_r}^{\alpha} G_{J_r}^{\beta} H_{J_r}^{\chi/\Delta}, \nonumber
 \end{align}
with $g_t^{\alpha\beta}=G_t^\alpha G_t^\beta$. One sees that,  when $ t_{ij}^{\alpha\beta} = t_{ij}  \delta_{\alpha\beta}$, the above Hamiltonian reduces to the classical formulas of the single-band $t-J$ model for cuprate superconductors \cite{FCZhang_1988}, where $g_t=G_t^2=\frac{2p}{1+p},\ g_J=G_J^2=\frac{4}{(1+p)^2}$, with doping $p=1-n$. In the single-band system, the physical superconducting order parameter is defined as $g_t|\Delta|$~\cite{FCZhang_1988}. In the multi-orbital system under consideration here, an immediate extension is made by employing the quantity $g_t^{\alpha\beta}|\Delta^{\alpha\beta}|$  to represent the $\alpha\beta-$orbital component of the superconducting order parameter. It is worth noting that at zero temperature $T=0$, approximate correspondence between the RMFT  and $U(1)$ slave boson mean-field theory (SBMFT) \cite{PhysRevB.38.5142,RevModPhys.78.17} self-consistent equations can be established if one assumes that $g_t^{ \alpha\beta}$ is related to the Bose condensation of holons, and $\Delta_{ }^{\alpha\beta}$ is linked to the spinon pairing in SBMFT.

Fourier transforms to Eq.~(\ref{eq:hjmf}-\ref{eq:mfij}), we obtain the mean-field Hamiltonian in momentum space

\begin{align}
\label{eq:mfk}
    &H^{\rm MF}=\sum_{\rm k}
    \Phi^{\dagger}_{\rm k}\left(
    \begin{array}{cc}
    H_{t,\rm k}^{\rm MF}+H_{J,\rm k}^{MF,\chi} & H_{J,\rm k}^{\rm MF,\Delta} \\
    {[} H_{J,\rm k}^{\rm MF,\Delta}{]}^\dagger & -{[}H_{t,\rm k}^{\rm MF}+H_{J,\rm k}^{\rm MF,\chi} {]}^* 
    \end{array}
    \right)\Phi_{\rm k}, 
\end{align}
where $\Phi_{\rm k}=(\Psi^T_{\rm{k}\uparrow},\Psi^\dagger_{-\rm k\downarrow})^T$ is the corresponding  Nambu  basis set. This equation can be solved  self-consistently,  combining  Eq.~(\ref{eq:chi}-\ref{eq:Delta}) to determine the final mean-field parameters  
, see also Supplementary Note 2 for more.


\begin{acknowledgments}
The authors thank the helpful discussions with Xunwu Hu, Zhong-Yi Xie, and Guang-Ming Zhang.
This project was supported by the National Key Research and Development Program of China (Grants No. 2022YFA1402802,
2018YFA0306001), the National Natural Science Foundation
of China (Grants No. 92165204, No.12174454, No. 11974432, No.12274472), the Guangdong Basic
and Applied Basic Research Foundation (Grants
No. 2022A1515011618, No. 2021B1515120015), Guangdong
Provincial Key Laboratory of Magnetoelectric
Physics and Devices (Grant No. 2022B1212010008),
Shenzhen International Quantum Academy (Grant No.
SIQA202102), and Leading Talent Program of Guangdong
Special Projects (201626003).
\end{acknowledgments}

\section*{Data availability}
The data is available in https://github.com/ZhihuiLuo/RMFT\_Ni327.

\bibliography{ref}

\section*{Author contributions}
Z.L. and B.L. contributed equally. D.X.Y. and W.W. conceived and designed the project. Z. L. wrote the code. Z.L. and B.L. performed the theoretical calculations and corresponding analysis under the supervision of D.X.Y. and W.W. All authors contributed to the interpretation of the results and wrote the paper.

\section*{Competing interests}
The authors declare no competing interests.
\end{document}


\preprint{APS/123-QED}
\title{Supplemental Information for: “High-T$_C$ superconductivity in La$_3$Ni$_2$O$_7$ based on the bilayer two-orbital t-J model”}
\date{\today}

\author{Zhihui Luo}
\thanks{These authors contributed equally to this work}

\author{Biao Lv}
\thanks{These authors contributed equally to this work}
\author{Meng Wang}
\author{W\'ei W\'u}
\email{wuwei69@mail.sysu.edu.cn}
\author{Dao-Xin Yao}
\email{yaodaox@mail.sysu.edu.cn}
\affiliation{Center for Neutron Science and Technology, Guangdong Provincial Key Laboratory of Magnetoelectric Physics and Devices, State Key Laboratory of Optoelectronic Materials and Technologies, School of Physics, Sun Yat-Sen University, Guangzhou, 510275, China}

\maketitle
\section{Model Details}
The tight-binding Hamiltonian $\mathcal{H}_t$ in Eq. (1) of the main text follows that from our previous works \cite{luo2023bilayer,wú2023charge}.  Here we rewrite it as,

\begin{eqnarray}
\mathcal{H}_t&=\sum_{\rm k \sigma} \Psi_{\rm k\sigma}^\dagger H(\rm k) \Psi_{\rm k\sigma}, 
\end{eqnarray}
with the matrix
\begin{eqnarray}
\label{eq:hk}
H(\rm{k})_{1,1}=H(\rm{k})_{3,3}&=2t_1^x (\cos k_x +\cos k_y)\nonumber \\&\quad+4t_2^x \cos k_x\cos k_y+\epsilon^x, \nonumber\\
H(\rm{k})_{2,2}=H(\rm{k})_{4,4}&=2t_1^z (\cos k_x +\cos k_y),\nonumber\\
&\quad+4t_2^z \cos k_x\cos k_y+\epsilon^z,\\
H(\rm{k})_{1,2}=H(\rm{k})_{3,4}&=2t^{xz}_3(\cos k_x-\cos k_y),\nonumber\\
H(\rm{k})_{1,4}=H(\rm{k})_{2,3}&=2t^{xz}_4(\cos k_x-\cos k_y),\nonumber \\
H(\rm{k})_{1,3}&=t^x_\bot, \nonumber\\
H(\rm{k})_{2,4}&=t^z_\bot. \nonumber
\end{eqnarray}
The basis is defined as $\Psi_\sigma=(c_{x_1\sigma},c_{z_1\sigma},c_{x_2\sigma},c_{z_2\sigma})^T$. The hopping parameters take the following values \cite{luo2023bilayer}: $t_1^x$=-0.483, $t_2^x$=0.069, $t_1^z$=-0.110, $t_2^z$=-0.017, $t_3^{xz}$=0.239, $t_4^{xz}$=-0.034, $t_\bot^x$=0.005, $t_\bot^z$=-0.635, and the site energies: $\epsilon^x$=0.776, $\epsilon^z$=0.409, in unit of eV.

\section{Details on RMFT}
We explicitly express the mean-field parameters in the main text of Eq.~(7)
\begin{widetext}
\begin{align}
    \chi_{\delta}^{\alpha\beta} =\frac{1}{N}\sum_{i}\langle c_{i\alpha\uparrow}^\dagger c_{i+\delta,\beta\uparrow}  \rangle=\frac{1}{N}\sum_{{\rm k},m}U_{\rm k}^*(\alpha,m)U_{\rm k}(\beta,m) e^{i{\rm k}\cdot \delta} f(\epsilon_{\rm k}^m),\\
\Delta_{\delta}^{\alpha\beta}=\frac{1}{N}\sum_{i}\langle c^\dagger_{i\alpha\uparrow} c^\dagger_{i+\delta,\beta\downarrow} \rangle =\frac{1}{N}\sum_{{\rm k},m}U_{\rm k}^*(\alpha,m)U_{\rm k}(\beta+4,m) e^{i{\rm k}\cdot \delta} f(\epsilon_{\rm k}^m).
\end{align}
\end{widetext}
Here $U_{\rm k}$ denotes the unitary transform  that diagonalizes Eq.~(7) in the main text, {\textit i.e.}, $U^\dagger_{\rm k} H^{\rm MF}_{\rm k} U_{\rm k}=D$ with $D$ the diagonal matrix, and $U_{\rm k}(\beta,m)$ the matrix element in $\beta$ row and $m$ column. It should be cautious in our case that $1\leq \alpha,\beta \leq 4$ while $1\leq m\leq 8$. Here $f(\epsilon_{\rm k}^m)=\frac{1}{e^{\epsilon_{\rm k}^m/k_{\rm B}T}+1}$ is the Fermi-Dirac function.
These two expressions are actually needed in a self-consistent  solution, combining the constraint on density
\begin{align}
\label{eq:constraint}
    n_{\alpha} =\frac{1}{N}\sum_{i}\langle c_{i\alpha\uparrow}^\dagger c_{i\alpha\uparrow}  \rangle=\frac{1}{N}\sum_{{\rm k},m}|U_{\rm k}(\alpha,m)|^2 f(\epsilon_{\rm k}^m)
\end{align}
to determine the saddle-point values. 
Such solution is mathematically equivalent to minimizing the free-energy $F=-k_{\rm B}T\ln \mathrm{Tr}[e^{-H/k_{\rm B}T}]$, which is expanded as
\begin{align}
    F&=F_1+F_2+F_3, \\
    F_1&=-k_{\rm B}T\sum_{{\rm k},m}\ln ( 2 \cosh \frac{|\epsilon_{\rm k}^{m}|}{2k_{\rm B}T}),\\
    F_2&=\sum_{{\rm k},\alpha} [H_{t,{\rm k}}^{\rm MF}+H_{J,{\rm k}}^{{\rm MF},\chi}]_{\alpha,\alpha},\\
    F_3&=\frac{3N}{4}\sum_{r,\delta} J_r G_J^\alpha G_J^\beta (|\chi_\delta^{\alpha\beta}|^2+|\Delta_\delta^{\alpha\beta}|^2).
\end{align}
Here the summation range of $\alpha,m$ are accordant with the previous, and $F_2$ indicates that only  the diagonal part of Eq.~(7) in the main text is required in the summation. 
By finding the minimal of $F$ under the constraint (\ref{eq:constraint}),
the saddle-point values of a series of  $\chi_\delta^{\alpha\beta},\Delta_{\delta}^{\alpha\beta}$ can be determined. We have verified that this solution agrees with the self-consistent solution.

Our above RMFT equations can be seen as from the finite temperature formalism established in Ref.~\cite{wang2010finite} in the large $U$ limit.
In Ref.~\cite{wang2010finite}, a $t-J-U$ model is utilized where an $U$ term is introduced  to allow finite double occupancy. 
In that work,  a correction term $-T\Delta S$ is introduced within the free-energy $F$ to incorporate the finite temperature effects on  Gutzwiller projection , which reads
\begin{align}\label{TDeltaS}
-T\Delta S = T\sum_i ( 
e_i \ln \frac{e_i}{e_{i0}} +q_i \ln \frac{q_i}{q_{i0}} + d_i \ln \frac{d_i}{d_{i0}}
).
\end{align}
Here $e_{i0},q_{i0},d_{i0}$ ($e_{i},q_{i},d_{i}$) represent empty, single and double occupancies at site $i$ before (after) projection, respectively.
These quantities are related by two conditions
\begin{align}
    e_{i0}+q_{i0}+d_{i0}=e_i+q_i+d_i=1,\label{eq:norm} \\
    q_{i0}+2d_{i0}=q_i+2d_i=n_i,\label{eq:n}
\end{align}
where Eq.~(\ref{eq:norm}) is the normalization condition and Eq.~(\ref{eq:n}) is imposed as the Gutzwiller projection does not change electron density. In the absence of magnetic ordering,  one finds~\cite{wang2010finite} $e_{i0}=(1-n_i/2)^2,q_{i0}=n_i(1-n_i/2)$ and $d_{i0}=n_i^2/4$ satisfying Eq.~(\ref{eq:norm}-\ref{eq:n}).
For given electron density $n_i$,  one hence in principle can minimize the $F-T\Delta S$ with respect to $e_i,q_i,d_i$ (and with respect to the mean-field parameters $\Delta, \chi$) to find the finite temperature solutions of the RMFT.
We note that due to the constraint in  Eq.~(\ref{eq:norm}-\ref{eq:n}), there is actually only one independent variable amongst $e_i,q_i,d_i$.
In our investigation,  as  the problem is considered in the large $U$ limit, \textit{i.e.,} $d_i=0$,  we further have  $e_i=1-n_i,q_i=n_i$.
Hence the correction term (\ref{TDeltaS}) becomes constant that only depends on the electron density $n_i$,
\begin{align}
    -T\Delta S = T\sum_i [
    (1-n_i)\ln (1-n_i) +(n_i-2) \ln (1-n_i/2)
    ].
\end{align}
Therefore, the finite temperature effects in Gutzwiller projection considered by Ref~\cite{wang2010finite} do not play a role in our study.  We note that  the $T_\mathrm{c}$ plot in the main text Fig.~1 for $n_x \approx 0.665 , n_z \approx 0.835$ corresponds to fairly large hole concentrations, where double occupancy and finite temperature effects on $g_t$ should be minimal. Moreover, in the slave boson mean-field theory (SBMFT) method~\cite{PhysRevB.38.5142}, which is closely related to RMFT~\cite{kotliar1986prl}, the superconducting $T_\mathrm{c}$ is  primarily  determined by the mean-field order parameter $\Delta>0$ at large hole concentrations.  In SBMFT, the holon  condensation,  which is analogous to $g_t$ in RMFT,  only affects $T_\mathrm{c}$ at small hole concentrations~\cite{PhysRevB.38.5142}. 

\section{BENCHMARK} 
To verify our calculations, we present a benchmark of 
 our RMFT results for one-band $t-J$ model. In Supplementary Fig.~\ref{fig:benchmark}, mean-field order parameters are shown as a function of $p$. The lines represents our calculations while the markers show results  from Ref.~\cite{FCZhang_1988}, with $J/t=0.2$, $T=0$. As can be seen,  they are in excellent agreement.
 
\begin{figure}[t]
	\centering
	\includegraphics[width=0.9\columnwidth]{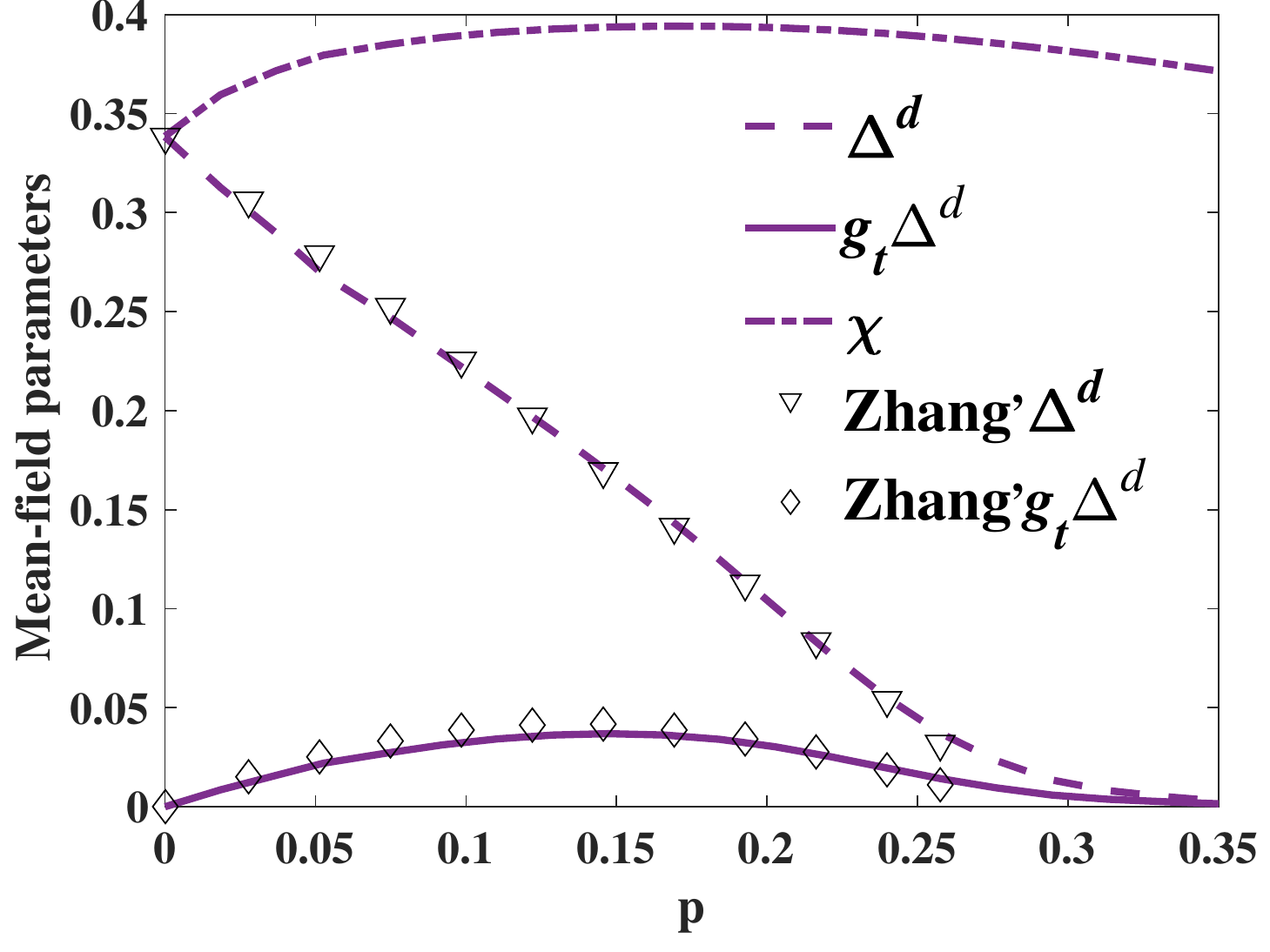}
	\caption{RMFT calculated mean-field order parameters as a function of $p$ for one-band $t-J$ model. The lines are from our calculations and markers are from Zhang's results \cite{FCZhang_1988}. $J/t=0.2$, $T=0$ are applied.}
	\label{fig:benchmark}
\end{figure}

\section{Doping phase diagram of $\Delta$}

\begin{figure}
	\centering
	\includegraphics[width=0.9\columnwidth]{Fig8.pdf}
	\caption{Intensities of different pairing gap parameters $|\Delta^\nu_{ \ell}|$ with vary doping $p_x,p_z$. The renormalization factor $g_t$ here is removed compared with Fig.~4 in the main text. A factor of 0.25 is incorporated in (c) for clarity. Also, a much smaller intensity range is used in the bottom panel.   $J_\bot=2J_{ ||}=0.18$ are applied.}
	\label{fig:phasediagram3}
\end{figure}

Supplementary Fig.~\ref{fig:phasediagram3} shows the magnitudes of six major pairing gap parameters $|\Delta^\nu_{ \ell}|$ with varying $p_x,p_z$. Compared with Fig.~4 in the main text, the 
renormalization factor $g_t$ is removed from the plot.  As can be seen the intensity distributions is roughly consistent with the former. The major difference appears in the lower boundary in Supplementary Fig.~\ref{fig:phasediagram3}a and left boundary in Supplementary Fig.~\ref{fig:phasediagram3}c, in which the magnitudes are notably enhanced.  This is expected since $g_t\rightarrow0$ as approaching half-filling. In physics, it reflects the suppression of the itinerancy of electrons by strong correlation effect, thereby inhibiting the formation of pairing bonds. 
Moreover, the stripe-like shape of intensity distributions in Supplementary Fig.~\ref{fig:phasediagram3}(a-b) seem overall move downward compared with Fig.~4(a-b) in the main text, which means, in other words, the optimal doping regime is more deviated from PC case if $g_t$ is ignored. 
In Supplementary Fig.~\ref{fig:phasediagram3}(e-f)  the  $s^\pm-$wave pairing fields $\Delta$ for $d_{x^2-y^2}$ orbital are also shown, which have smaller magnitudes compared with the $d_{z^2}$ counterparts.

\section*{Supplementary Reference}
\bibliography{ref}